\documentclass[]{spie}  

\usepackage{amsmath,amsfonts,amssymb}
\usepackage{graphicx}
\usepackage[colorlinks=true, allcolors=blue]{hyperref}
\usepackage{iftex}
\usepackage{lipsum}
\usepackage{mdframed}
\usepackage{soul}
\usepackage{multirow}
\usepackage[font=normal]{caption}
\ifTUTeX
  \usepackage{fontspec}
\else
  \usepackage[T1]{fontenc}
  \usepackage[utf8]{inputenc} 
  \DeclareUnicodeCharacter{200B}{{\hskip 0pt}}
\fi
\title{The European Low Frequency Survey on the Simons Array}

\author[a,b]{Aniello Mennella}
\author[c]{Kam Arnold}
\author[d]{Susanna Azzoni}
\author[e]{Carlo Baccigalupi}
\author[f]{A. J. Banday}
\author[g]{Rita Bel\'en Barreiro}
\author[h]{Darcy Barron}
\author[a,b]{Marco Bersanelli}
\author[g]{Francisco J. Casas}
\author[c]{Sean Casey}
\author[i,j]{Elena de la Hoz}
\author[a,b]{Cristian Franceschet}
\author[k]{Michael E. Jones}
\author[l]{Ricardo T. Gen\'ova-Santos}
\author[l]{R. Hoyland}
\author[i]{Adrian T. Lee}
\author[g]{Enrique Martinez-Gonzalez}
\author[a]{Filippo Montonati}
\author[l]{José-Alberto Rubi\~no-Martín}
\author[k]{Angela Taylor}
\author[g]{Patricio Vielva}

\affil[a]{University of Milan, Department of Physics, Via Celoria 16, Milan, Italy}
\affil[b]{INFN Milan section, Via Celoria 16, Milan, Italy}
\affil[c]{University of California at San Diego, 5998 Alcala Park, San Diego, CA 92110-8001, USA}
\affil[d]{Princeton University, Department of Physics, Princeton, NJ 08542-0430, USA}
\affil[e]{SISSA; Via Bonomea 265,34136 Trieste, Italy}
\affil[f]{IRAP, Universit\'e de Toulouse, CNRS, CNES, UPS, (Toulouse), France}
\affil[g]{IFCA-CSIC, Avenida de los Castros, 39005 Santander, Cantabria, Spain}
\affil[h]{The University of New Mexico, Dept. of Physics and Astronomy, Albuquerque, NM 87131, USA}
\affil[i]{University of Berkeley, Department of Physics, 366 Physics North MC 7300 Berkeley, CA, 94720-7300, USA}
\affil[j]{CNRS-UCB International Research Laboratory, Centre Pierre Binétruy, IRL2007, CPB-IN2P3, Berkeley, CA 94720, USA}
\affil[k]{Oxford University, Department of Physics, Parks Road, Oxford, OX1 3PU, UK}
\affil[l]{IAC, Calle Vía Láctea s/n, E-38205 La Laguna, Tenerife, Spain}

\authorinfo{Further author information: \\
Aniello mennella: aniello.mennella@fisica.unimi.it\\  
José-Alberto Rubi\~no-Martín: jalberto@iac.es\\
Michael Jones: Mike.Jones@physics.ox.ac.uk\\
Adrian Lee: adrian.lee@berkeley.edu}

\begin{document} 
\maketitle

\begin{abstract}
In this paper we present the European Low Frequency Survey (ELFS), a project that will enable foregrounds-free measurements of the primordial $B$-mode polarization and a detection of the tensor-to-scalar ratio, $r$, to a level $\sigma(r) = 0.001$ by measuring the Galactic and extra-galactic emissions in the 5--120\,GHz frequency window. Indeed, the main difficulty in measuring the B-mode polarization comes from the fact that many other processes in the Universe also emit polarized microwaves, which obscure the faint Cosmic Microwave Background (CMB) signal. The first stage of this project is being carried out in synergy with the Simons Array (SA) collaboration, installing a 5.5--11\,GHz (X-band) coherent receiver at the focus of one of the three 3.5\,m SA telescopes in Atacama, Chile, followed by the installation of the QUIJOTE-MFI2 in the 10--20 GHz range. We designate this initial iteration of the ELFS program as ELFS-SA. The receivers are equipped with a fully digital back-end that will provide a frequency resolution of 1\,MHz across the band, allowing us to clean the scientific signal from unwanted radio frequency interference, particularly from low-Earth orbit satellite mega constellations. This paper reviews the scientific motivation for ELFS and its instrumental characteristics, and provides an update on the development of ELFS-SA.
\end{abstract}


\section{Introduction}
\label{sec:intro}  

Nowadays the leading contender to understand the initial conditions of the Big Bang is inflation. According to this paradigm, the primordial Universe was composed of a single quantum field, the inflaton, ​which caused a rapid, faster-than-light expansion that stretched fluctuations in the field from the quantum to macroscopic scales. 

Inflation can be tested, as it provides a unique mechanism to generate a primordial background of gravitational waves that must have left its
imprint in the CMB, an as-yet undetected signature in the CMB polarization, the so-called $B$-modes. This characteristic
pattern in the polarization is faint (less than a millionth of a degree in temperature) and the main difficulty comes from the fact that many other objects in the Universe also emit polarized
microwaves. 

The only way to separate the CMB from the foregrounds is to exploit the fact that all these
components change brightness with frequency in a different way. Our current knowledge of the synchrotron radiation has increased significantly after ground-based experiments operating in the 2--20\,GHz range: S-PASS
(2.3\,GHz) \cite{krachmalnicoff2018} in the Southern hemisphere, C-BASS (5\,GHz) \cite{jones2018} and QUIJOTE-MFI (10--20\,GHz) \cite{MFIwidesurvey} in the Northern hemisphere. These data highlight that the synchrotron emission is more complex than has been assumed in CMB forecast codes. Data from low-frequency instruments will be key in constraining and removing the synchrotron emission to extract the CMB signal.

The European Low Frequency Survey (ELFS) is a long-term plan to deploy dedicated telescopes to produce a full-sky survey in the 5--100\,GHz range with an  angular resolution of $\sim$20 arcmin at 10\,GHz, sub-GHz spectral resolution and sensitivity that will allow $B$-mode extraction from data produced by current and future CMB experiments. ELFS-SA is the combination of a 5.5--11\,GHz (X-band) receiver and the 10--20\,GHz QUIJOTE-MFI2\cite{hoyland2022}. Both receiver will be installed in sequence in the Gregorian focus of one of the Simons Array telescopes \cite{barron2018}. 

In this paper we discuss the design and main instrumental characteristic of the X-band receiver and show the impact of ELFS-SA measurements when combined with those expected from the Simons Observatory.
​

\section{The X-band receiver}
\label{sec_xband_receiver}

\begin{figure} [h!]
\begin{center}
    \begin{tabular}{c} 
    \includegraphics[height=5cm]{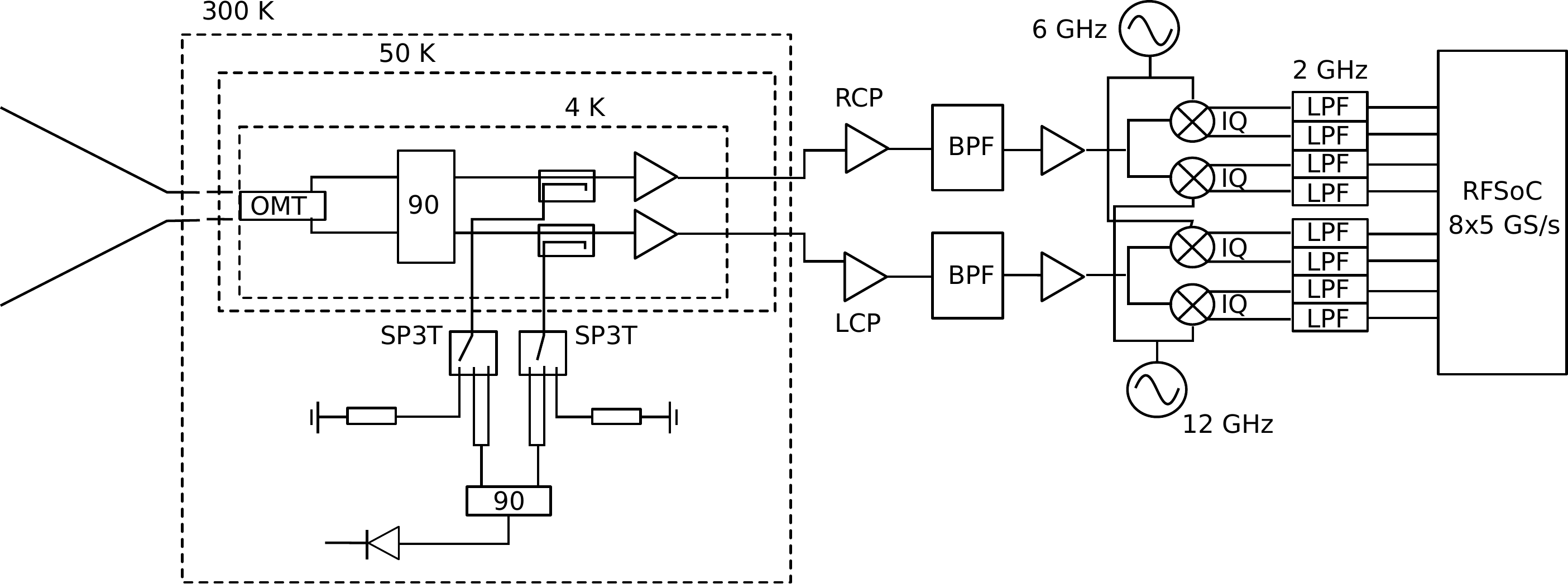}
    \end{tabular}
\end{center}
\caption{\label{fig_receiver_schematics}
        Schematic diagram of the ELFS-SA X-band receiver.}
\end{figure}

Figure~\ref{fig_receiver_schematics} shows a schematic view of the X-band receiver. The front-end will be cooled to 4\,K in a cryostat adapted from the C-BASS North receiver \cite{2014MNRAS.438.2426K}. This is based on a Sumitomo SRDK-408D2 two-stage Gifford-McMahon cold head that can be interfaced up to four low-noise amplifiers plus the associated planar hybrid modules to allow for the continuous comparison radiometer architecture used in C-BASS. It will be modified to accommodate the new 2:1 bandwidth OMT in place of the 30\,\% bandwidth orthomode transducer (OMT) used in C-BASS.

The sky signal reflected by the telescope propagates through a room-temperature corrugated feedhorn. An OMT at 4\,K splits the signal into two perpendicularly polarized waves that are subsequently combined by a 90$^\circ$ hybrid to produce left- and right-hand circularly polarized signals. Two Low Noise Amplifiers (LNAs), model LNF\_LNC4\_16C from Low Noise factory, provide the primary gain, while coaxial 30\,dB couplers in the RF lines before the LNAs can inject a noise signal for calibration.

The back-end uses an FPGA-based digital unit and shares the same design of the QUIJOTE-MFI2 instrument \cite{hoyland2022}. The MFI2 FPGA (Xilinx ZCU208 Ultrascale) can simultaneously acquire eight RF channels at a sampling frequency of 5.0\,GSps, with a $2.5$\,GHz band and 1\,MHz spectral resolution. The back-end will divide the full bandwidth into spectral sub-bands with maximum bandwidth of 2.5\,GHz, which are down-converted to base band $[0,2.5]$\,GHz through separate Local Oscillators (LOs). In our design we will achieve the down-conversion by using the complex output of mixers to obtain two bands from each LO. 

The backend design implements a hard programmed polyphase filterbank and a fast Fourier transform to retrieve the spectral information from the digitised samples of the fast onboard ADCs in the time domain. The power spectral density is then integrated in time and averaged. A temporary storage space is used to store 24\,h of raw data used to find an optimum blocking filter for any undesired interference. The final stored scientific signal is a spectral average with an averaging factor that depends on the scientific needs and the available storage space.

The OMT inherits from a design used for the Square Kilometre Array Band 5a and 5b feeds. This is a quad-ridge design in which the ridges for one polarization are held at a constant spacing while they pass by the coaxial probe and backshort of the other polarization. This design reduces coupling between the polarizations and allows for a bandwidth of more than 2:1 with minimal excitation of higher order modes. The OMT is manufactured in four quadrants which are assembled with precision dowels to set the critical spacings between the ridges, which then requires no further tuning.  

\begin{figure} [h!]
\begin{center}
    \includegraphics[height=5cm]{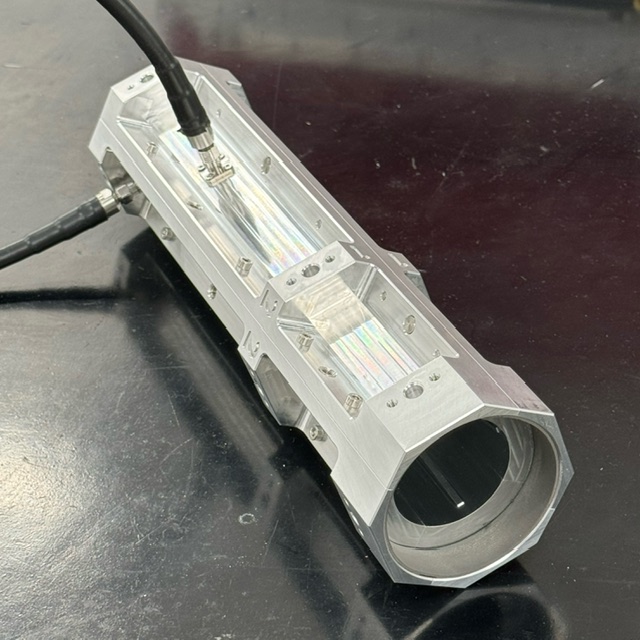}\\
    \begin{tabular}{cc} 
    \includegraphics[width=8cm]{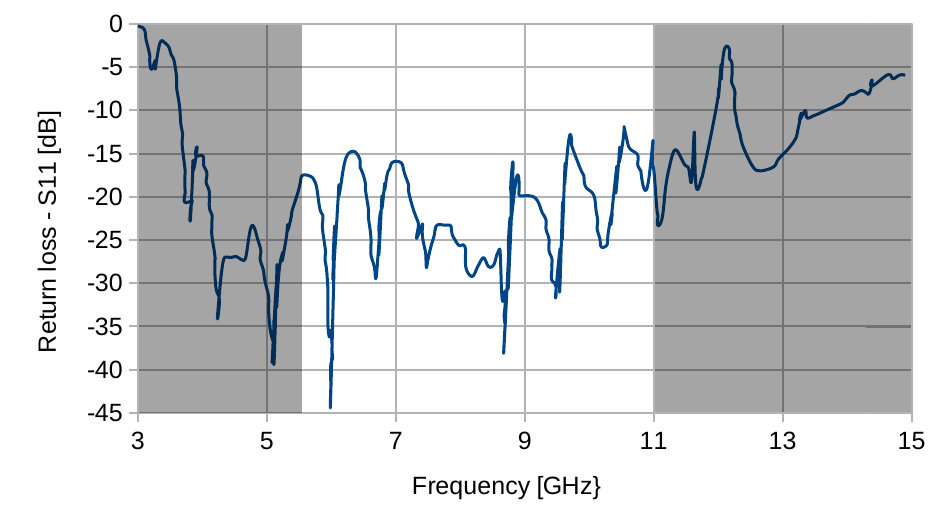}&
    \includegraphics[width=8cm]{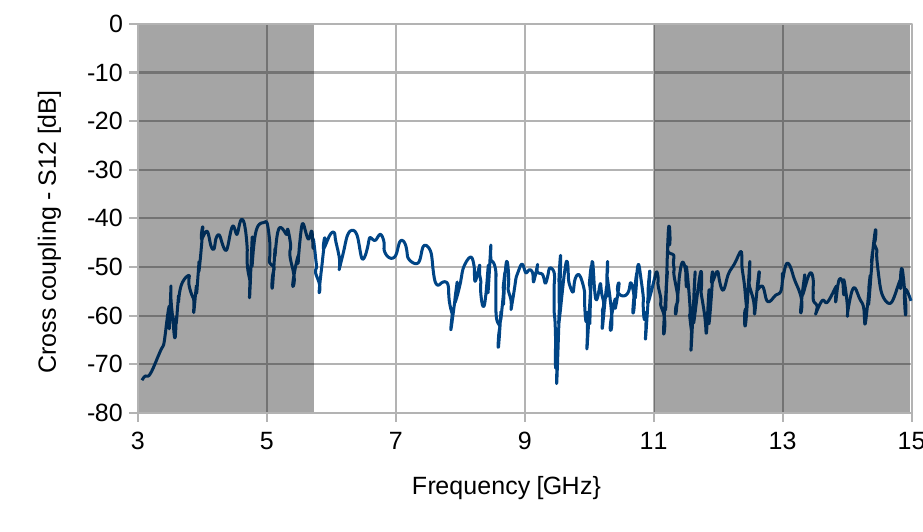}\\
    \includegraphics[width=8cm]{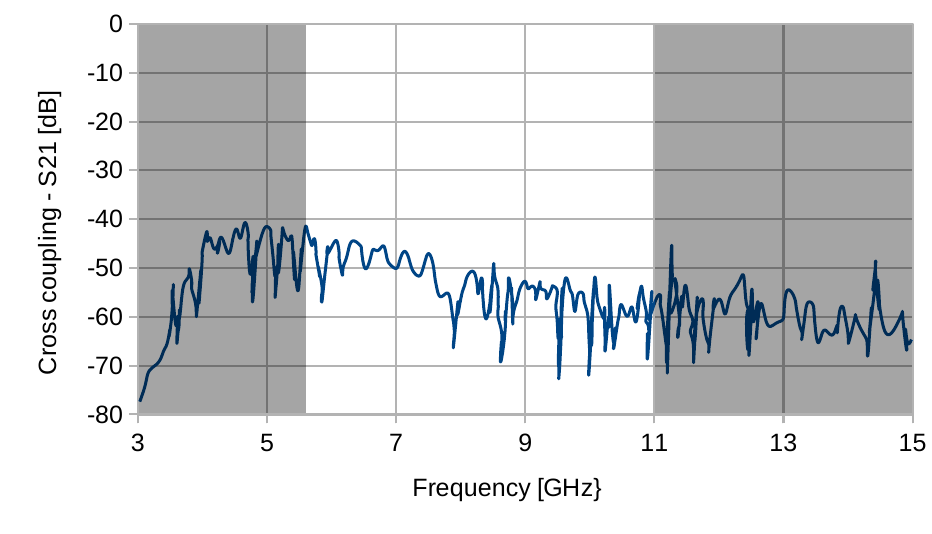}&
    \includegraphics[width=8cm]{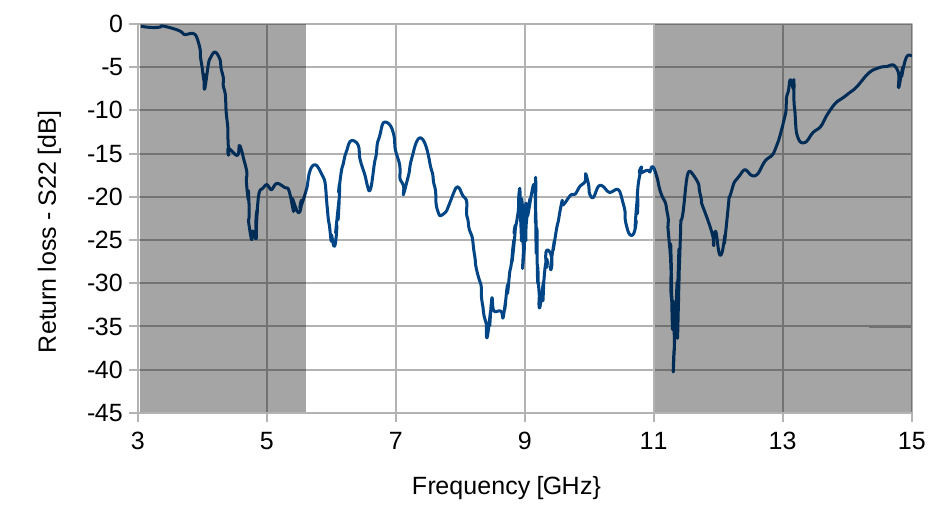}
    \end{tabular}
\end{center}
\caption{\label{fig_omt}
        The ELFS-SA OMT. Top picture: the fabricated OMT showing the ouput coaxial ports. Bottom panel: the measured OMT performance. Top-left and bottom-right: the return loss at the output ports. Top-right and bottom-left: cross-coupling between the output ports. The white area highlights the working band.}
\end{figure}

The top panel of Fig.~\ref{fig_omt} shows the manufactured OMT with the two coaxial cable outputs. The panels below the picture show the measured return loss and cross-coupling at the two output ports. We see that in the working band (highlighted by the white area) the return loss is generally better than $-15$\,dB and the coupling less than $-40$\,dB. 

\begin{figure} [h!]
\begin{center}
    \includegraphics[width=16cm]{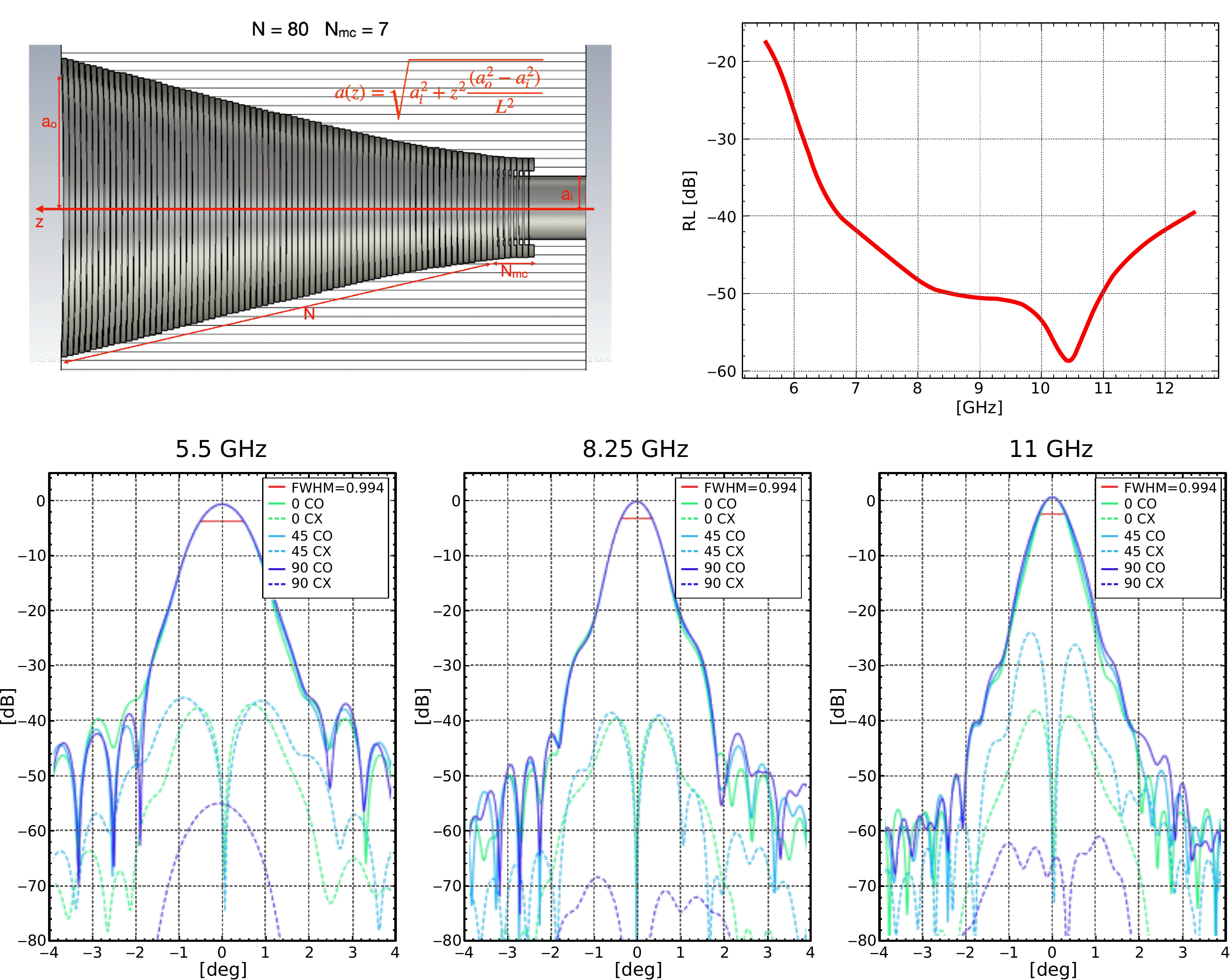}
\end{center}
\caption{\label{fig_feed}
        The ELFS-SA X-band feedhorn. Top-left: the feedhorn design. Top-right: simulated return loss. Bottom panels: the simulate co-polar and cross-polar beams.}
\end{figure}

The top-left panel of Fig.~\ref{fig_feed} shows the feedhorn electromagnetic design with the main dimensions and parameters. We based the design on the ideas presented in Granet and James [\citenum{1487785}], choosing a hyperbolic profile and a ring-loaded slot mode converter. The corrugation teeth are 4.5\,mm in the body of the horn and 5\,mm in the mode converter part, with constant tooth/groove ratio of 0.889. This configuration allowed us to obtain excellent broadband performance in terms of return loss and cross-polarization performance, as shown in Fig.\,\ref{fig_feed}.

The return loss, in particular, is better than $-20$\,dB and for a large fraction of the band (at frequencies $\gtrsim 6.5$\,GHz) it is less than $-40$\,dB. The beam is highly symmetric, with a maximum cross-polarization of the order of $-40$\,dB, apart from the high edge of the band where it increases to $-30$\,dB. The sidelobes are less than $-50$\,dB across all the band.

\begin{figure} [h!]
\begin{center}
    \includegraphics[width=\textwidth]{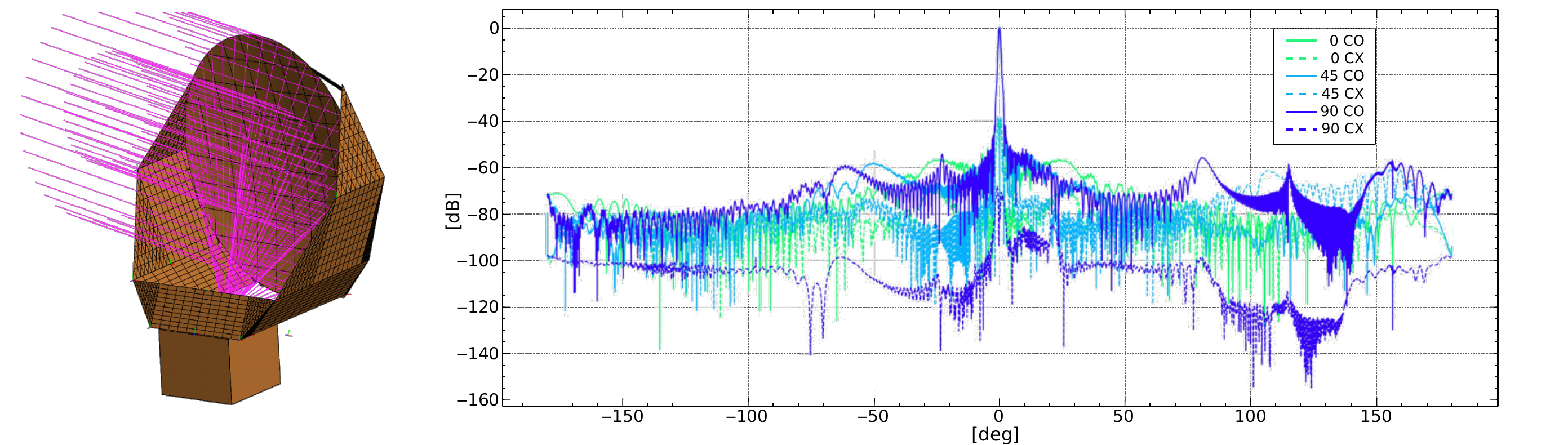}
\end{center}
\caption{\label{fig_feed-telescope}
        The ELFS-SA X-band optical system performance. Left: ray-tracing of the feedhorn at the focus of the SA telescope. Right: simulated beam pattern of the entire optical system.}
\end{figure}

The left panel of Figure~\ref{fig_feed-telescope} shows the Grasp model of the Simons Array telescope, which we have simulated considering the feedhorn beam pattern at its focus. The right panel shows the result of a preliminary simulation, performed neglecting the presence of the baffling structure. We see that also in this worst-case scenario the far sidelobes sit at $-40$\,dB maximum and the cross-polar response never exceeds $-50$\,dB.

\section{Impact on science}
\label{sec_impact_on_science}

To assess the impact of adding low frequency channels in foreground and CMB data analysis, we considered the Simons Observatory\footnote{We include the additional SATs from SO:UK and SO:JP as well as an extended observation time (until 2035).} (SO) experiment combined with the ELFS-SA channels (the X-band and the QUIJOTE-MFI2 instruments). 

In particular we focus on the ability of removing the foreground signals in presence of a synchrotron emission with various level of complexity and in presence of limited knowledge about the sky. We first describe the simulated instrumental configurations, then we present the input sky models used in the simulations and, finally, we show how the addition of spectral information at low frequencies improves the possibility to detect a complex spectral behavior in the synchrotron emission.

\subsection{Instrument characteristics}
    Table~\ref{tab_instrument_sensitivities} summarizes the main instrumental parameters used in our simulations. 
    We used the noise parameters to generate $100$ noise simulations for each instrument, following the same methodology as detailed in Ade et al [\citenum{Ade_2019}] and Wolz et al [\citenum{so_forecast_2023}]. We model the noise power spectra for each band of each instrument as the sum of two noise sources:
        \begin{equation}
            N_{\ell} = N_{\rm white}\left[ 1 + \left(\dfrac{\ell}{\ell_{\rm knee}}\right)^{\alpha_{\rm knee}} \right] \, ,
            \label{eq:red_noise_ps}
        \end{equation}
    i.e., a white noise component, represented by a constant $N_\text{white}$, plus a $1/f$ component modeled as a power law at the power spectrum level. This $1/f$ component models the noise that arises from the atmosphere and electronic noise, and it is characterized by an exponent denoted as $\alpha_\text{knee}$ and a ``knee'' multipole at $\ell_\text{knee}$.
    
\begin{figure}[h!]
        \begin{center}
            \includegraphics[width=0.6\textwidth]{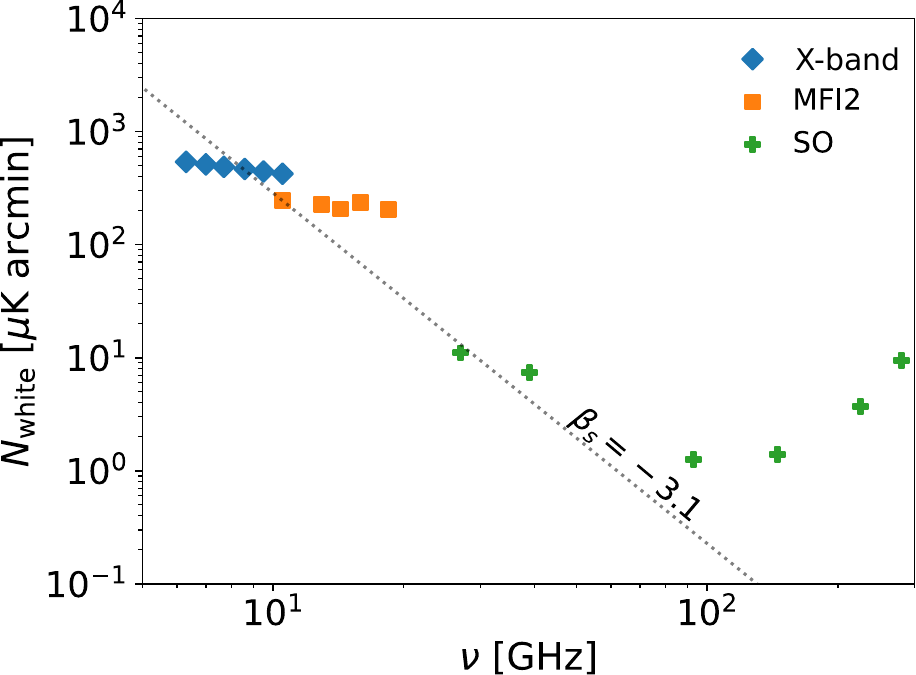}
        \end{center}
        \caption{\label{fig_instrument_sensitivities}Instrument sensitivities as a function of frequency. The dotted line shows the extrapolated sensitivity following a power law with an exponent of $-3.1$.}
    \end{figure}

    Figure~\ref{fig_instrument_sensitivities} shows the sensitivity vs. frequency for the various instruments. The dotted line shows the extrapolated sensitivity assuming a spectral index $\beta = -3.1$.
    
    \begin{table}[h!]
        \caption{\label{tab_instrument_sensitivities} Instrument and noise specifications}
        \begin{center}
        \begin{tabular}{cccccc}
            \hline
            Experiment   &  $\nu$ & FWHM  &  N$_{\rm white}$  & $\ell_{\rm{knee}}$ & $\alpha_{\rm{knee}}$\\
                        & [GHz]  & [arcmin] & [$\mu$K$\cdot$arcmin] &             &\\
            \hline
            \multirow{5}{*}{Nominal} & 27 & 91    & 33    & 15   & $-$2.4 \\
            & 39    & 63    & 22    & 15    & $-$2.4 \\
            \multirow{3}{*}{SO-SAT}& 93    & 30    & 2.5   & 25    & $-$2.5 \\
            & 145   & 17    & 2.8   & 25    & $-$3   \\
            & 225   & 11    & 5.5   & 35    & $-$3   \\
            & 280   & 9     & 14    & 40   & $-$3   \\
            \hline
            \multirow{6}{*}{X-band} & 6.3 & 46.6  & 539    & 15    & $-$2.4 \\
            & 7   &  42.2     & 512    & 15    & $-$2.4  \\
            & 7.7   & 38.1    & 487    & 15    & $-$2.4 \\
            & 8.6   & 34.4    & 465   & 15    & $-$2.4 \\
            & 9.5   & 31.1    & 443   & 15    & $-$2.4 \\
            & 10.5  & 28.1    & 423   & 15    & $-$2.4 \\
            \hline
            \multirow{5}{*}{MFI2} & 10.5 & 33.7  & 245    & 15    & $-$2.4 \\
            & 12.9   &  27.4   & 228    & 15    & $-$2.4  \\
            & 14.3   & 24.8    & 206    & 15    & $-$2.4 \\
            & 15.9   & 22.2    & 236   & 15    & $-$2.4 \\
            & 18.4   & 19.2    & 203   & 15    & $-$2.4 \\
            \hline
        \end{tabular}
        \end{center}
    \end{table}

\subsection{Simulated sky components}

    In our simulations we considered three components: (i) the CMB, (iii) the synchrotron emission, and (ii) the thermal dust emission (see Table~\ref{tab_sky_characteristics}).
    
    For the CMB we simulated 100 maps derived from power spectra compatible with \textit{Planck} best-fit cosmological parameters \cite{planck_cosmo_par} without tensor fluctuations ($r=0$) and including lensing, i.e. the non-Gaussian $B$-mode pattern arising from gravitational lensing of the CMB photons. Furthermore, we explored an alternative scenario in which the lensing effect is partially removed, thus producing another set of $100$ CMB maps that incorporate a $50$\,\% delensing procedure as described in Namikawa et al [\citenum{so_delensing}].

    Regarding the synchrotron emission we considered four cases. We based the first and second cases on the \texttt{PySM}\footnote{https://github.com/galsci/pysm.}\cite{PYSM2} \texttt{s5} and \texttt{s7} models, which assume a simple power law and a power law with negative curvature (i.e. a spectral index dependent on frequency) with a pivot frequency of $23$\,GHz. The third and fourth cases, that we name \texttt{sC2} and \texttt{sC}, are compatible with current statistics obtained from the analysis of the Southern hemisphere sky. In the \texttt{sC2} template the curvature is compatible with a Gaussian distribution $\mathcal{N}\left(0.04, 0.1\right)$ \cite{krachmalnicoff2018}. The \texttt{sC} template was obtained by renormalizing the \texttt{PySM} \texttt{s7} curvature template to a Gaussian distribution with a mean of 0.04 and a standard deviation of 0.1. 
    
    Equation~(\ref{eq_synchrotron_equation}) summarizes the analytical models describing the polarized synchrotron emission, where $Q_\text{s}$ and $U_\text{s}$ are the linear polarization Stokes parameters, $a_\text{s}^Q$ and $a_\text{s}^U$ are the amplitude terms, $\beta_\text{s}$ is the spectral index and $c_\text{s}$ the curvature.
    
    \begin{equation}
        \begin{array}{ll}
            \text{Power\,law}&
            \begin{pmatrix}
                Q_\text{s} \\
                U_\text{s}
            \end{pmatrix}_{\nu} =
            \begin{pmatrix}
                a_\text{s}^Q \\
                a_\text{s}^U
            \end{pmatrix}
            \displaystyle\left(\frac{\nu}{30\,\mathrm{GHz}}\right)^{\beta_\text{s}}\\
            \\
            \text{Power\,law\,with\,curvature}&
            \begin{pmatrix}
                Q_\text{s} \\
                U_\text{s}
            \end{pmatrix}_{\nu} =
            \begin{pmatrix}
                a_\text{s}^Q \\
                a_\text{s}^U
            \end{pmatrix}
            \displaystyle\left(\frac{\nu}{30\,\mathrm{GHz}}\right)^{\beta_\text{s} + c_\text{s} \log\left(\frac{\nu}{23 \mathrm{GHz}}\right)}
        \end{array}
        \label{eq_synchrotron_equation}
    \end{equation}

    Finally, we adopted the \texttt{PySM} \texttt{d10} model to simulate the thermal dust emission. This model assumes a simple spectrum with the shape of a modified black-body with pixel-dependent parameters (see Eq.~(\ref{eq_dust_sed})). Although recent analyses show that  the dust emission might be more complex (see, for example, Ritacco et al [\citenum{dust_ritacco}]), we adopted the simplest model as our focus is on the synchrotron emission.
    
    In Eq.~(\ref{eq_dust_sed}) $Q_\text{d}$ and $U_\text{d}$ are the linear polarization Stokes parameters, $a_\text{d}^Q$ and $a_\text{d}^U$ are the amplitude terms, $\beta_\text{d}$ is the spectral index and $T_\text{d}$ the dust temperature.
    
    \begin{equation}
        \begin{pmatrix}
            Q_\text{d} \\
            U_\text{d}
        \end{pmatrix}_{\nu}
        =
        \begin{pmatrix}
            a_\text{d}^Q \\
            a_\text{d}^U
        \end{pmatrix}
        \left(\frac{\nu}{353\,\mathrm{GHz}}\right)^{\beta_\text{d}-2}\frac{B(\nu, T)}{B(353\,\text{GHz}, T_\text{d})}     
        \label{eq_dust_sed}
    \end{equation}

    \begin{table}[h!]
    \caption{\label{tab_sky_characteristics}Sky components included in the simulation}
        \begin{center}
            \begin{tabular}{ccl}
            \hline
            Component   & code & Characteristics \\
            \hline
            \multirow{2}{*}{CMB} & \texttt{cL} & Lensed CMB ($a_L=1$). \\
            & \texttt{cDL} & Delensed CMB ($a_L=0.5$)$^1$. \\
            \hline
            \multirow{3}{*}{Synchrotron} & \texttt{s5} & PySM \texttt{s5} model (power law). \\
             & \texttt{s7} & PySM \texttt{s7} model (power law+curvature). \\
            & \texttt{sC2} & \multirow{2}{*}{Custom models adding a curvature term.}\\
            & \texttt{sC} & \\
            \hline
            Thermal dust & \texttt{d10} & PySM \texttt{d10} model (modified black-body). \\
            \hline
            \\
            \multicolumn{3}{l}{$^1$ In this scenario we assume that 50\% of the lensing effect can be}\\
            \multicolumn{3}{l}{\phantom{$^1$} removed. Hence the parameter $a_L=0.5$.}
            \end{tabular}
        \end{center}
    \end{table}
    
\begin{figure}[h!]
        \begin{center}
            \includegraphics[width=0.7\textwidth]{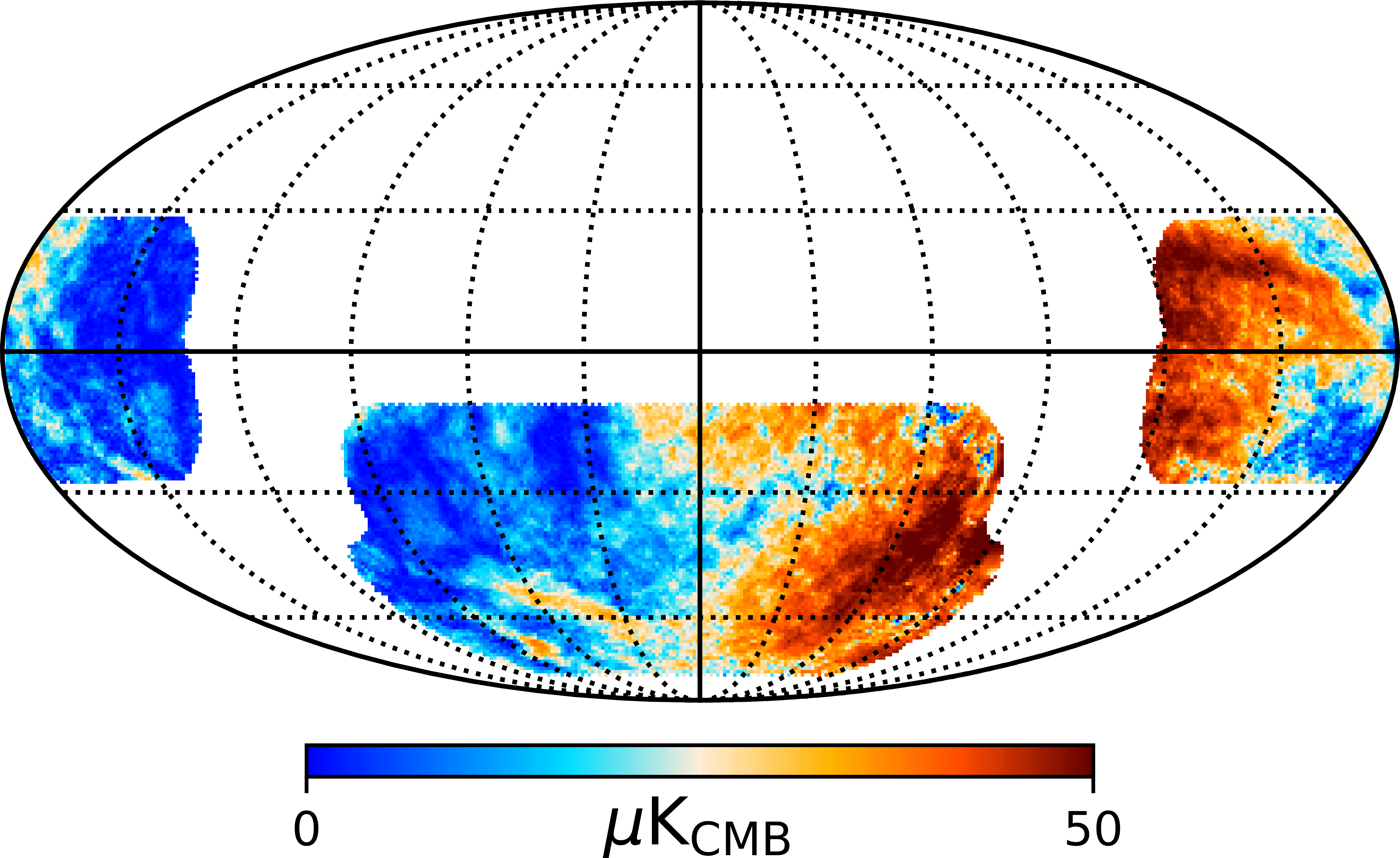}
        \end{center}
        \caption{\label{fig_synchrotron_sky}Polarized synchrotron amplitude intensity at 27 GHz.}
    \end{figure}

    Figure~\ref{fig_synchrotron_sky} shows a map of the polarized synchrotron amplitude at 27\,GHz (\texttt{s5} model) in the sky patch expected to be observed by SO.

\subsection{Component separation and fitting procedure}

    To separate the various signals we apply the component separation algorithm \texttt{B-SeCRET}\cite{de2020detection}. This is a parametric pixel-based, maximum-likelihood method, which relies on an affine-invariant Markov chain Monte Carlo ensemble sampler (\texttt{emcee}\footnote{\url{https://emcee.readthedocs.io/en/stable/}}) to draw samples from a posterior distribution\cite{foreman2013emcee}. Before applying \texttt{B-SeCRET} we convolve the maps with a uniform Gaussian beam with a FWHM of $91$\,arcmin, and then we downgrade them to the HEALPix\cite{gorski_2005} parameter $N_{\rm side} = 64$.

    \texttt{B-SeCRET} applies Bayesian inference to determine the best-fit model parameters given some prior information, which is crucial to reduce parameter degeneracy like, for example, the degeneracy between $\beta_\text{s}$ and $c_\text{s}$.
    Instead of sampling the full posterior distribution we sample the conditional probability distributions of amplitude and spectral parameters separately. This strategy effectively diminishes the dimensionality of the problem, ultimately helping convergence. The interested reader can find more information on this approach in de la Hoz et al [\citenum{de2020detection}].
    
\subsection{Results}

    In this section we assess the impact of adding the ELFS-SA frequencies to the SO channels in the ability to detect a synchrotron spectral behavior that could be more complex than that assumed in component separation. To this aim we applied component separation to the various simulated skies assuming various models and assessed how the quality of the fit changes when the assumed model does not match the input one.

\begin{figure}[h!]
        \begin{center}
         \includegraphics[width=0.7\textwidth]{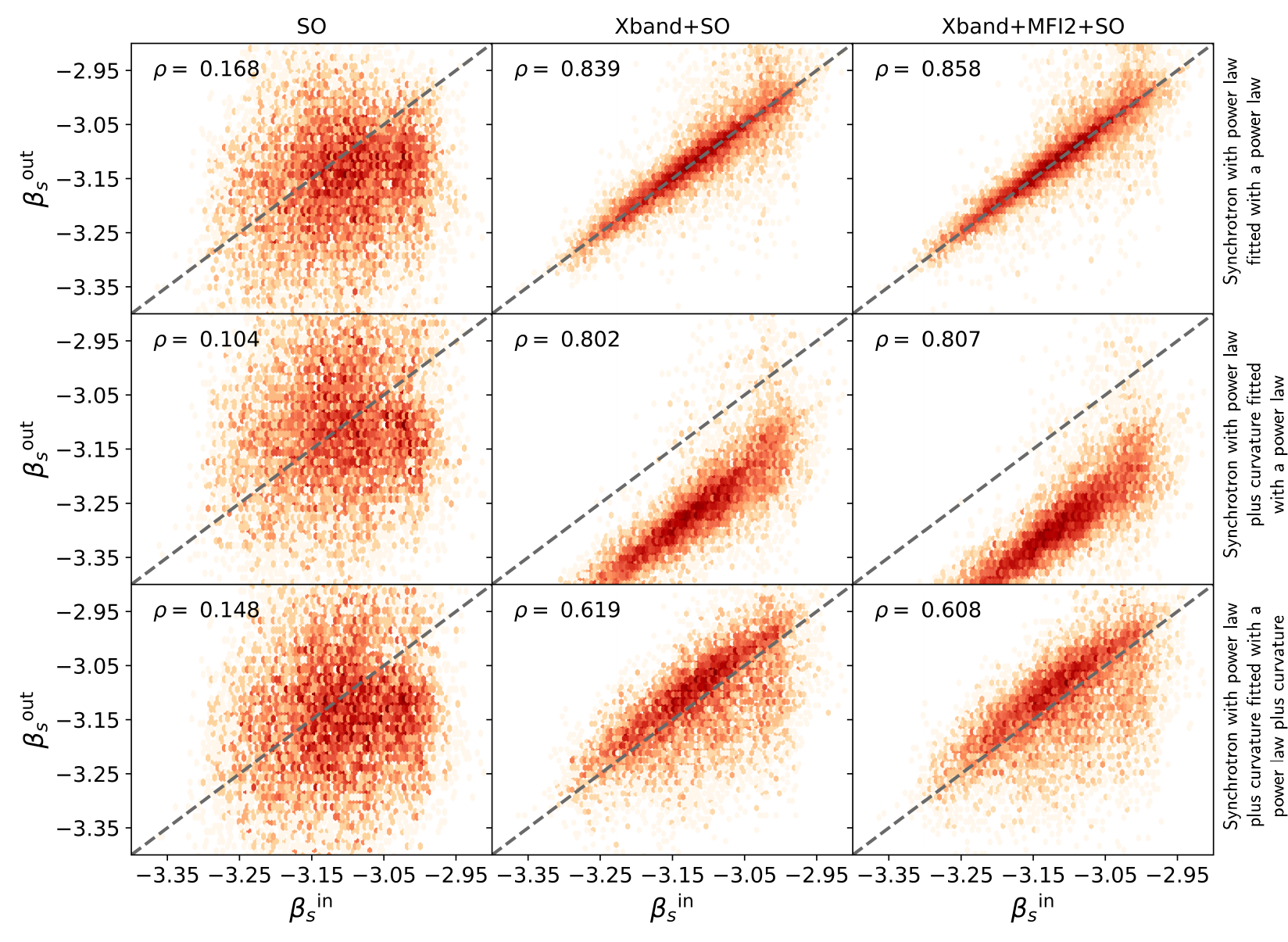}
         \caption{\label{fig_synchrotron_fits}Comparison of recovered $\beta_\text{s}$ values to input $\beta_\text{s}$. From left to right: SO, X-band+SO, and X-band+MFI2+SO results. The dashed diagonal represents $\beta_\text{s}^\mathrm{out} = \beta_\text{s}^\mathrm{in}$. The correlation coefficients ($\rho$) are displayed in the upper left of each plot.}
        \end{center}
    \end{figure}

    Figure~\ref{fig_synchrotron_fits} displays the fitted synchrotron spectral index for various instrument configurations. The first row shows the case where synchrotron follows a power-law and it is fitted with a power-law. The second row represents the scenario where synchrotron has a curved spectral index (model \texttt{sC}) but is fitted with a power-law. Lastly, the third row demonstrates the case where curved synchrotron is fitted with the correct model. 
    
    This result clearly shows that without the low frequency channels the algorithm is not able to spot whether the synchrotron spectral behavior is more complex than the assumed model. The inclusion of ELFS-SA, instead, generates a bias in the case when the input and assumed skies do not match, allowing one to detect the presence of a curvature in the synchrotron emission spectral index.
    
\begin{figure}[h!]
        \begin{center}
         \includegraphics[width=\textwidth]{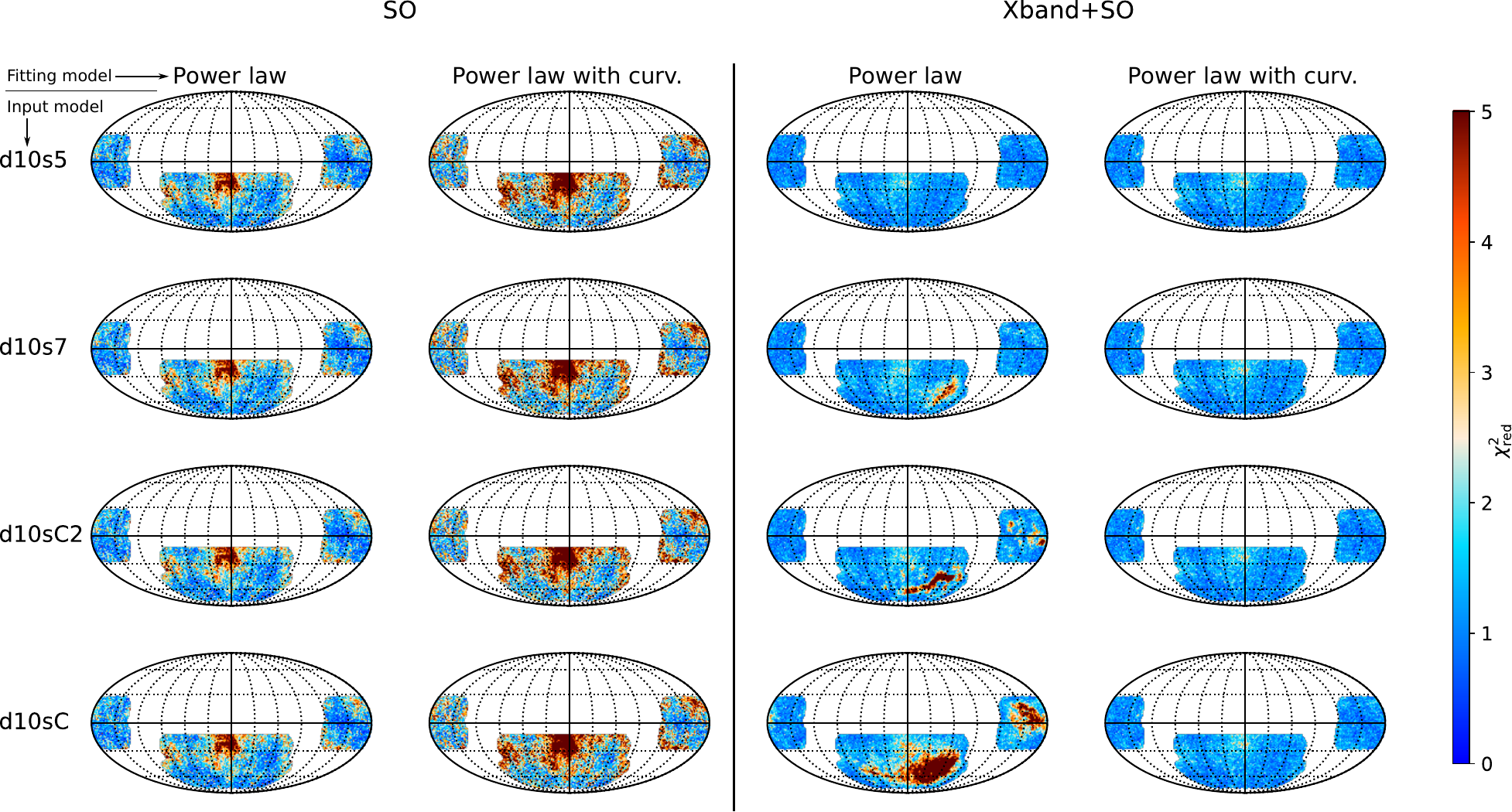}
         \caption{\label{fig_chi2_maps}Reduced $\chi^2$ maps generated using SO data alone and ELFS+SO data. The various rows correspond to simulations performed using the different input sky models. The odd columns display the maps obtained with a power law model fitting the synchrotron emission, while the even columns present the $\chi^2_\text{red}$ values when the synchrotron is modeled with a power law with curvature.}
        \end{center}
    \end{figure}

    Figure~\ref{fig_chi2_maps} shows maps of the reduced chi square, $\chi^2_\text{red}$, obtained when the input sky is fitted either with a simple power law (odd columns) or with a model including curvature (even columns). The two columns on the left correspond to results obtained with SO alone, while the two columns on the right represent results obtained adding ELFS-SA channels.
        
    If we compare the left with the right columns we see that SO alone lacks the discerning power necessary to distinguish between the various synchrotron models. The introduction of an additional parameter, such as curvature, does not improve the fits, because the lack of channels below 30\,GHz limits the SO constraining power. The fact that the recovered values align with the actual values depends essentially on the accuracy of the prior information. If the prior expected value deviated significantly from the true value, the analysis results would be less favorable. Finally, the lack of constraining power implies that the results tend to favor simplest model, regardless of the synchrotron actual complexity, which could lead to elevated foreground residuals and substantial biases in the recovered tensor-to-scalar ratio.

    On the contrary, if we add the ELFS-SA channels we systematically obtain better results when we add curvature in the model. In the case where we fit a curvature model to a sky containing a simple power-law synchrotron (last two maps in the first row) we obtain essentially no difference in the $\chi^2_\text{red}$ maps, as we should expect, given that the simple power-law is a particular case of a model with curvature with $c_\text{s} = 0$.

\section{Conclusions}
\label{sec_conclusions}

In this paper we presented ELFS, a plan to deploy dedicated telescopes to produce a full-sky survey in the 5--100\,GHz range, and its first incarnation, ELFS-SA, which foresees the deployment of a 5.5--11\,GHz receiver in the Gregorian focus of one of the Simons Array telescopes, followed by the installation of the QUIJOTE-MFI2 to cover the 10--20\,GHz range. 

We analyzed the potential of ELFS-SA in discerning complex synchrotron behavior, when combined with measurements from next generation experiments, like the Simons Observatory (SO). We have shown that with all the considered synchrotron models the inclusion of additional low-frequency bands significantly reduces  the foreground residuals. We expect that this reduction will be particularly critical when SO achieves its optimal performance, as the absence of low-frequency information could lead to a biased  detection of the tensor-to-scalar ratio if foreground residuals are not effectively addressed.

In the next steps we will extend our study to the recovery of the tensor-to-scalar ratio parameter and apply the analysis to other CMB experiments such as LiteBIRD and CMB-S4.

\section*{Acknowledgements}

This is not an official SO Collaboration paper.

\bibliography{report} 
\bibliographystyle{spiebib} 

\end{document}